\documentclass[twocolumn,english,aps,prl,amsmath,amssymb,nofootinbib,showpacs,superscriptaddress]{revtex4}
\usepackage{amsmath}
\usepackage{amssymb}
\usepackage{graphicx}
\usepackage{wasysym}
\usepackage{esint}
\usepackage{pdfpages}

\makeatletter

\newcommand*\LyXThinSpace{\,\hspace{0pt}}

\@ifundefined{textcolor}{}
{%
 \definecolor{BLACK}{gray}{0}
 \definecolor{WHITE}{gray}{1}
 \definecolor{RED}{rgb}{1,0,0}
 \definecolor{GREEN}{rgb}{0,1,0}
 \definecolor{BLUE}{rgb}{0,0,1}
 \definecolor{CYAN}{cmyk}{1,0,0,0}
 \definecolor{MAGENTA}{cmyk}{0,1,0,0}
 \definecolor{YELLOW}{cmyk}{0,0,1,0}
}

\usepackage[english]{babel}
\usepackage{latexsym}
\usepackage{graphics}
\usepackage{epsfig}
\usepackage{color}
\usepackage{bm}

\usepackage{babel}

\usepackage{babel}

\makeatother

\usepackage{babel}
\begin{document}

\title{Dimensional crossover in a strongly interacting ultracold atomic
Fermi gas}

\author{Umberto Toniolo}

\affiliation{Centre for Quantum and Optical Science, Swinburne University of Technology,
Melbourne 3122, Australia}

\author{Brendan C. Mulkerin}

\affiliation{Centre for Quantum and Optical Science, Swinburne University of Technology,
Melbourne 3122, Australia}

\author{Chris J. Vale}

\affiliation{Centre for Quantum and Optical Science, Swinburne University of Technology,
Melbourne 3122, Australia}

\author{Xia-Ji Liu}

\affiliation{Centre for Quantum and Optical Science, Swinburne University of Technology,
Melbourne 3122, Australia}

\affiliation{Kavli Institute for Theoretical Physics, UC Santa Barbara, USA}

\author{Hui Hu}

\affiliation{Centre for Quantum and Optical Science, Swinburne University of Technology,
Melbourne 3122, Australia}

\date{\today}
\begin{abstract}
We theoretically explore the crossover from three dimensions (3D)
to two (2D) in a strongly interacting atomic Fermi superfluid through
confining the transverse spatial dimension. Using the gaussian pair
fluctuation theory, we determine the zero-temperature equation of
state and Landau critical velocity as functions of the spatial extent
of the transverse dimension and interaction strength. In the presence
of strong interactions, we map out a dimensional crossover diagram
from the location of maximum critical velocity, which exhibits distinct
dependence on the transverse dimension from 2D to quasi-2D, and to
3D. We calculate the dynamic structure factor to characterize the
low-energy excitations of the system and propose that the intermediate
quasi-2D regime can be experimentally probed using Bragg spectroscopy. 
\end{abstract}


\maketitle
Recent breakthroughs in understanding strongly interacting ultracold
atomic Fermi gases at the crossover from Bose-Einstein condensates
(BEC) to Bardeen-Cooper-Schrieffer (BCS) superfluids \cite{Eagles1969,Leggett1980,NSR1985,SadeMelo1993}
have attracted enormous attention from diverse fields of physics \cite{MBX1999,Giorgini2008,Randeria2014}.
Due to the unprecedented accuracy in controlling the dimensionality
and interatomic interaction \cite{Bloch2008,Chin2010}, significant
progress has been made to realize systems in the 2D limit \cite{Martiyanov2010,Feld2011,Frohlich2011,Dyke2011,Orel2011,Koschorreck2012,Sommer2012,Zhang2012,Makhalov2014,Ong2015,Ries2015,Murthy2015,Dyke2016,Martiyanov2016,Fenech2016,Boettcher2016,Cheng2016}.
It thus provides a new paradigm to explore a number of intriguing
low-dimensional phenomena, including the absence of a true long-range
order at nonzero temperature \cite{Mermin1966,Hohenberg1967}, the
existence of quasi-condensates due to the Berezinskii-Kosterlitz-Thouless
mechanism \cite{Berezinskii1972,KT1973,Salasnich2013}, the disruptive
role of pair fluctuations around the mean-field (MF) \cite{Randeria1989,SchmittRink1989,Engelbrecht1990,Watanabe2013,Bauer2014,Marsiglio2015,He2015,Mulkerin2015,Bighin2016},
and the possible observation of exotic imbalanced superfluidity \cite{Conduit2008,Yin2014,Toniolo2017}.
These unusual features lie at the heart of many technologically interesting
materials such as high-temperature superconductors \cite{Lee2006},
where the dimensional crossover from 3D to 2D is dictated by the ratio
of the Cooper pair size to the thickness of the superconducting
layer.

Despite rapid experimental advances, the fundamental criteria for reaching
the strict 2D regime at the BEC-BCS crossover are still not well understood.
Experimentally, a 2D Fermi gas is realized by freezing the atomic
motion in the transverse direction using a single highly-oblate harmonic
trap \cite{Dyke2011,Fenech2016} or a tight one-dimensional optical
lattice \cite{Martiyanov2010,Feld2011,Sommer2012,Zhang2012,Ries2015}.
In the absence of interactions, the 2D condition is easy
to clarify within the single-particle picture: the chemical potential
$\mu$ and temperature $k_{B}T$ of the system should be smaller than
the characteristic energy scale $\hbar\omega_{z}$ along the transverse
direction, so that all atoms stay in the lowest transverse mode \cite{Dyke2011}.
With strong interactions, the situation is less clear. Indeed, a recent
measurement of time-of-flight expansion indicates that it is difficult
to display the strict 2D kinematics when the interaction becomes stronger
\cite{Dyke2016}. Theoretically, the
dimensional crossover of a strongly interacting Fermi gas from 3D
to 2D is challenging due to the strong correlations~\cite{Levinsen2015}. %
To date, an interacting quasi-2D
Fermi gas has only been studied in the highly imbalanced polaron
limit \cite{Levinsen2012} or by using mean-field approach that is
known to break down in the 2D limit \cite{Hu2011,Fischer2014}.

\begin{figure}
\begin{centering}
\includegraphics[width=0.48\textwidth]{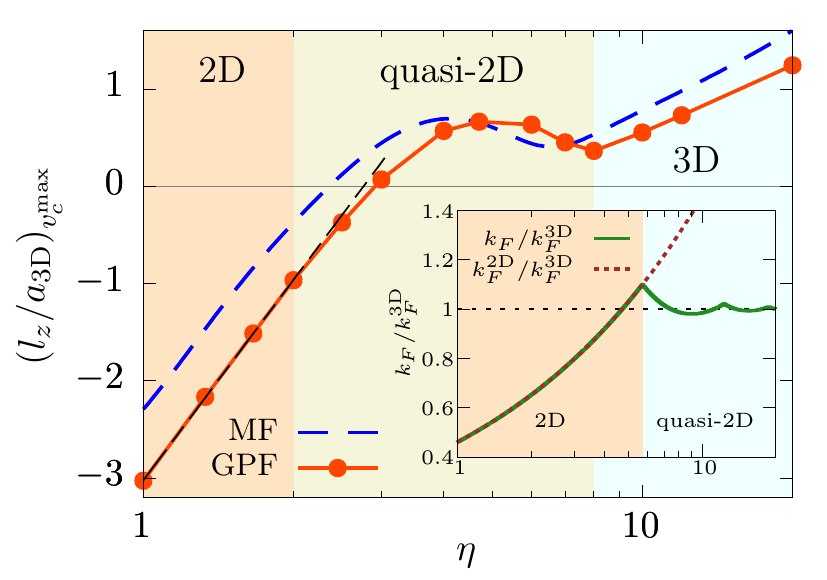} 
\par\end{centering}
\caption{(color online). The dimensional crossover diagram, tuned by the dimension
parameter $\eta$ (in logarithmic scale) and the value of the interaction
strength, $(l_{z}/a_{\text{3D}})_{v_{c}^{\max}}$, at which the Landau
critical velocity peaks. The red solid line (with circles) and blue
dashed line show $(l_{z}/a_{\text{3D}})_{v_{c}^{\max}}$ predicted
by the GPF and MF theories, respectively. Their distinct dependences
on $\eta$ enables us to identify the 2D and 3D regimes, and the quasi-2D
regime in between. The inset shows the dimensional crossover diagram
in the non-interacting case, determined from the free Fermi gas number
equation (see text). \label{fig:phase_diagram}}
\end{figure}

In this Letter, we determine the dimensional
crossover diagram (see Fig. \ref{fig:phase_diagram}), by considering
a \emph{uniform} strongly interacting quasi-2D Fermi gas with periodic
boundary condition (PBC) in the tightly confined transverse direction.
This configuration is motivated by the recent successful production
of a box trapping potential that leads to a uniform Bose or Fermi
gas in bulk \cite{Gaunt2013,Mukherjee2016}. We apply %
a gaussian pair fluctuation (GPF) theory to obtain the %
zero-temperature equation of state (Fig. \ref{fig:EoS})~\cite{Mulkerin2015}
and Landau critical velocity (Fig. \ref{fig:Landau_critical_velocity})
at the dimensional crossover. For a given dimensional parameter $\eta\equiv k_{F}^{\text{3D}}l_{z}$,
where $l_{z}$ is the periodic length of the confining potential in
the transverse direction and $k_{F}^{\text{3D}}\equiv(3\pi^{2}n)^{1/3}$
is the three-dimensional (3D) Fermi momentum of the gas with density
$n$, we determine the interaction strength at which there is a maximum
of the Landau critical velocity~\cite{footnote}, $(l_{z}/a_{\text{3D}})_{v_{c}^{\max}}$,
where $a_{{\rm 3D}}$ is the 3D $s$-wave scattering length. A %
Fermi superfluid is most robust to external excitations at this maximum, which is found, in 3D, close to unitarity~\cite{Combescot2006,Sensarma2006,Miller2007}.
We obtain a regime where the maximum of the critical velocity in the
BEC-BCS crossover depends on the logarithm of $\eta$ for $\eta<2$,
denoting a 2D regime (i.e., the long-dashed line in Fig. \ref{fig:phase_diagram}).
Also, $(l_{z}/a_{\text{3D}})_{v_{c}^{\max}}$ depends linearly on
$\eta$ for $\eta>8$, denoting a 3D regime. The region that links
these regimes is defined as quasi-2D and has properties distinct to
the 2D and 3D limits. 
%

\textit{Theoretical framework.} \textemdash{} We start by defining
various Fermi momenta. We consider a $s$-wave
two-component Fermi gas at zero temperature where the transverse direction
is confined with periodic length $l_{z}$, implying the discretization
of momentum in the $z$-direction, $k_{z}=2\pi n_{z}/l_{z}$, for
any integer $n_{z}$. The Fermi momentum $k_{F}$ of the dimensional
crossover system can then be defined as the maximally allowed momentum
in the axial direction: 
\begin{equation}
n=\frac{1}{2\pi l_{z}}\sum_{n_{z}=-n_{\max}}^{n_{\max}}\left[k_{F}^{2}-\left(\frac{2\pi n_{z}}{l_{z}}\right)^{2}\right],\label{eq:kF}
\end{equation}
where $n_{\max}$ is the largest integer smaller than $k_{F}l_{z}/(2\pi)$.
It is useful to first examine the dimensional crossover diagram for
an ideal Fermi gas, as shown in the inset of Fig.~\ref{fig:phase_diagram}.
At large $l_{z}$ (or $\eta$), $k_{F}$ approaches the 3D Fermi momentum
$k_{F}^{\textrm{3D}}$, as anticipated. In the limit of small $l_{z}$,
instead, $k_{F}$ coincides with a 2D Fermi momentum $k_{F}^{\textrm{2D}}\equiv\sqrt{2\pi n_{\textrm{2D}}}=\sqrt{2\eta/(3\pi)}k_{F}^{\textrm{3D}}$,
where the column density $n_{\textrm{2D}}\equiv nl_{z}$. An ideal
2D Fermi gas is thus realized when $k_{F}=k_{F}^{\textrm{2D}}$ or
$\eta<\sqrt[3]{6}\pi\simeq5.7$, for which only the lowest transverse
mode is occupied. This simple 2D condition is not applicable in the
presence of strong interactions, a situation that we shall consider
below. %
A strongly interacting Fermi gas with contact interactions between
unlike fermions can be described by a single-channel Hamiltonian density
\cite{Randeria1989,He2015,Hu2006,Diener2008}, 
\begin{alignat}{1}
\mathcal{H}=\sum_{\sigma=\uparrow,\downarrow}\bar{\psi}_{\sigma}(\mathbf{r})\mathcal{H}_{0}\psi_{\sigma}(\mathbf{r})-g\bar{\psi}_{\uparrow}(\mathbf{r})\bar{\psi}_{\downarrow}(\mathbf{r})\psi_{\downarrow}(\mathbf{r})\psi_{\uparrow}(\mathbf{r}),\label{eq:Hamiltonian}
\end{alignat}
where $\psi_{\sigma}(\mathbf{r})$ are the annihilation operators
for each spin state, $\mathcal{H}_{0}=-\hbar^{2}\nabla^{2}/(2M)-\mu$
is the free Hamiltonian with atomic mass $M$, $\mu$ is the chemical
potential, and $g>0$ denotes the bare interaction strength. The contact
potential is a convenient choice of interaction, however it needs
to be regularized and related to a physical observable of the system.
We achieve this by relating the bare interaction strength $g$ to
the bound state energy $B_{0}$ \cite{Yamashita2014}, 
\begin{equation}
\frac{1}{g}=\sum_{\mathbf{k},k_{z}}\frac{1}{2\left(\epsilon_{\mathbf{k}}+\epsilon_{k_{z}}\right)+B_{0}},\label{eq:interaction_regularization}
\end{equation}
%
where $\epsilon_{\mathbf{k}}=\hbar^{2}\mathbf{k}^{2}/(2M)$ and the sums on $(\mathbf{k},k_z)$ carry a volume %
factor that goes to $(2\pi)^2l_z$ at the thermodynamic limit. In order
to recover the 3D limit, we require the two-body $T$-matrix in the
dimensional crossover be equivalent to its 3D counterpart in the limit
$l_{z}\rightarrow\infty$. This implies that the binding energy, $B_{0}$,
can be analytically related to the 3D scattering length $a_{\text{3D}}$,
according to \cite{Petrov2001,Yamashita2014}, 
\begin{equation}
B_{0}=4\left(\frac{\hbar^{2}}{Ml_{z}^{2}}\right)\textrm{arcsinh}^{2}\left[\frac{e^{l_{z}/\left(2a_{\textrm{3D}}\right)}}{2}\right].\label{eq:binding_energy}
\end{equation}
It is also possible to define a 2D binding energy, $\epsilon_{B}^{\text{2D}}\equiv\hbar^{2}/(Ma_{\text{2D}}^{2})$,
find the equivalence between the scattering $T$-matrix and the 2D
$T$-matrix as $l_{z}\rightarrow0$, and show analytically that $B_{0}=\epsilon_{B}^{\text{2D}}$
in the 2D limit.

We solve the many-body Hamiltonian Eq. \eqref{eq:Hamiltonian} by
using the zero-temperature GPF theory, which provides reasonable quantitative %
predictions for equation of state in both 2D \cite{He2015}
and 3D \cite{Hu2007,Hu2010}. The theory takes into account strong
pair fluctuations at the gaussian level on top of mean-field solutions
\cite{Hu2006,Diener2008} and hence we separate the thermodynamic
potential into two parts: $\Omega=\Omega_{{\rm MF}}+\Omega_{{\rm GF}}$.
The mean-field part is~\cite{He2015}, 
\begin{equation}
\Omega_{\textrm{MF}}=\frac{\Delta^{2}}{g}+\sum_{\mathbf{k},k_{z}}\left(\xi_{\mathbf{k},k_{z}}-E_{\mathbf{k},k_{z}}\right),\label{eq:OmegaMF}
\end{equation}
where $\smash{\xi_{\mathbf{k},k_{z}}=\epsilon_{\mathbf{k}}+\epsilon_{k_{z}}-\mu}$,
$\smash{E_{\mathbf{k},k_{z}}=\sqrt{\xi_{\mathbf{k},k_{z}}^{2}+\Delta^{2}}}$,
and the order parameter $\Delta$ is determined self-consistently
using the mean-field gap equation, $\Delta\sum_{\mathbf{k},k_{z}}[(\epsilon_{\mathbf{k}}+\epsilon_{k_{z}}+B_{0}/2)^{-1}-E_{\mathbf{k},k_{z}}^{-1}]=0$,
ensuring the gapless Goldstone mode \cite{Combescot2006}. The pair
fluctuation part is given by ($Q\equiv(\mathbf{q},q_{z},\omega)$)
\cite{He2015}, 
\begin{equation}
\Omega_{\text{GF}}=\sum_{\mathbf{q},q_{z}}\varint_{0}^{\infty}\frac{d\omega}{2\pi}\ln\left[\frac{\mathbf{M}_{11}\left(Q\right)\mathbf{M}_{11}\left(-Q\right)-\mathbf{M}_{12}^{2}\left(Q\right)}{\mathbf{M}_{11}^{C}\left(Q\right)\mathbf{M}_{11}^{C}\left(-Q\right)}\right],\label{eq: OmegaGF}
\end{equation}
with the matrix elements, 
\begin{eqnarray}
\mathbf{M}_{11}\left(Q\right) & = & \frac{1}{g}+\sum_{\mathbf{k},k_{z}}\left(\frac{u_{+}^{2}u_{-}^{2}}{\omega-E_{+}-E_{-}}-\frac{v_{+}^{2}v_{-}^{2}}{\omega+E_{+}+E_{-}}\right),\nonumber \\
\mathbf{M}_{12}\left(Q\right) & = & \sum_{\mathbf{k},k_{z}}\left(-\frac{u_{+}u_{-}v_{+}v_{-}}{\omega-E_{+}-E_{-}}+\frac{u_{+}u_{-}v_{+}v_{-}}{\omega+E_{+}+E_{-}}\right),\nonumber \\
\mathbf{M}_{11}^{C}\left(Q\right) & = & \frac{1}{g}+\sum_{\mathbf{k},k_{z}}\frac{u_{+}^{2}u_{-}^{2}}{\omega-E_{+}-E_{-}}.
\end{eqnarray}
Here, we use the notations $E_{\pm}\equiv E_{\mathbf{k}\pm\mathbf{q}/2,k_{z}\pm q_{z}/2}$,
$u_{\pm}^{2}=(1+\xi_{\mathbf{k}\pm\mathbf{q}/2,k_{z}\pm q_{z}/2}/E_{\mathbf{k}\pm\mathbf{q}/2,k_{z}\pm q_{z}/2})/2$
and $v_{\pm}^{2}=1-u_{\pm}^{2}$ \cite{He2015,Hu2006,Diener2008}.
The chemical potential is found by solving the number equation, $n=-\partial\Omega/\partial\mu$.

\begin{figure}
\centering{}\centering \includegraphics[width=0.41\textwidth]{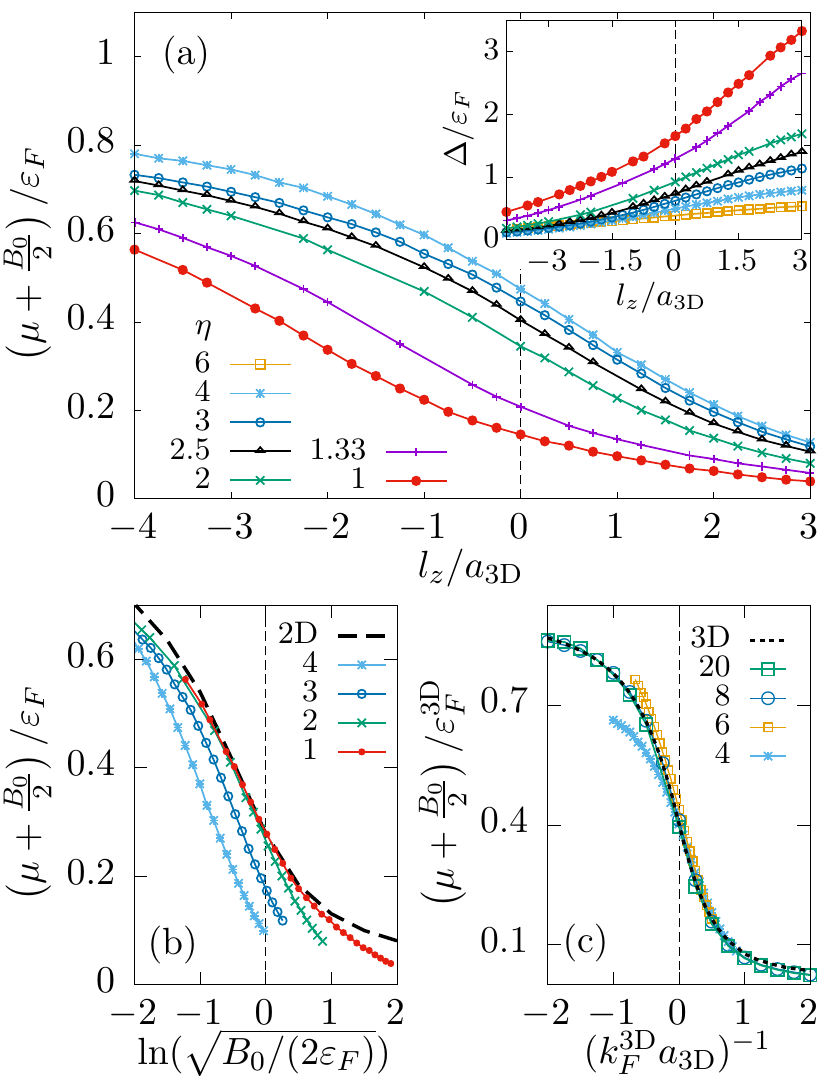}
\caption{(color online). (a) The dimensionless shifted chemical potential,
$(\mu+B_{0}/2)/\varepsilon_{{\rm F}}$, as a function of $l_{z}/a_{\text{3D}}$
at various dimension parameters ($\eta=1\sim6$), from the quasi-2D
to 2D regime. The inset shows the order parameter $\Delta/\varepsilon_{{\rm F}}$.
(b) The chemical potential near the 2D limit, replotted as a function
of $\ln(\sqrt{B_{0}/(2\varepsilon_{F})})$. (c) The chemical potential
near the 3D limit in units of $\varepsilon_{F}^{\text{3D}}$ is shown
as a function of $1/(k_{F}^{\text{3D}}a_{\text{3D}})$. \label{fig:EoS}}
\end{figure}

\textit{Equation of state.} \textemdash{} In Fig. \ref{fig:EoS}(a),
we report the dimensionless shifted chemical potential $(\mu+B_{0}/2)/\varepsilon_{F}$,
where $\varepsilon_{F}=\hbar^{2}k_{F}^{2}/(2M)$ is the Fermi energy,
at the BEC-BCS crossover tuned by $l_{z}/a_{\text{3D}}$ and at the
dimensional crossover tuned by $\eta=1\sim6$. For all values of $\eta$,
the dependence of the chemical potential on $\eta$ remains similar
to the typical decreasing slope found in 3D \cite{Hu2006,Diener2008}.
However, as $\eta$ decreases the curves shift towards negative values
of $l_{z}/a_{\text{3D}}$. The inset plots the order parameter, $\Delta/\varepsilon_{F}$,
and we see a similar behavior to the chemical potential as we decrease
$\eta$. As $\eta$ approaches the 2D limit, we can compare the magnitude
of the chemical potentials with the 2D case through the interaction
parameter $\ln(\sqrt{B_{0}/(2\varepsilon_{F})})$, as shown in Fig.
\ref{fig:EoS}(b). We plot a range of dimensions, $\eta=1\sim4$,
and the 2D result (black dashed), and see a clear trend of the chemical
potential approaching the 2D result for $\eta\lesssim 2$. %
In Fig. \ref{fig:EoS}(c), we compare the chemical potential
to the 3D result (black short dashed), where we plot the chemical
potential in units of the 3D Fermi energy $\varepsilon_{F}^{{\rm 3D}}$
as a function of $1/(k_{F}^{\text{3D}}a_{\text{3D}})$. We find %
excellent agreement in the BEC limit for $\eta\gtrsim4$
and by $\eta>8$ the dimensional crossover system is effectively
in the 3D limit for the entire BEC-BCS crossover. Thus, we see a
distinct quasi-2D regime for the dimension parameter $2<\eta<8$.
This observation is confirmed below by the calculation of Landau critical
velocity.

\begin{figure}
\centering \includegraphics[width=0.41\textwidth]{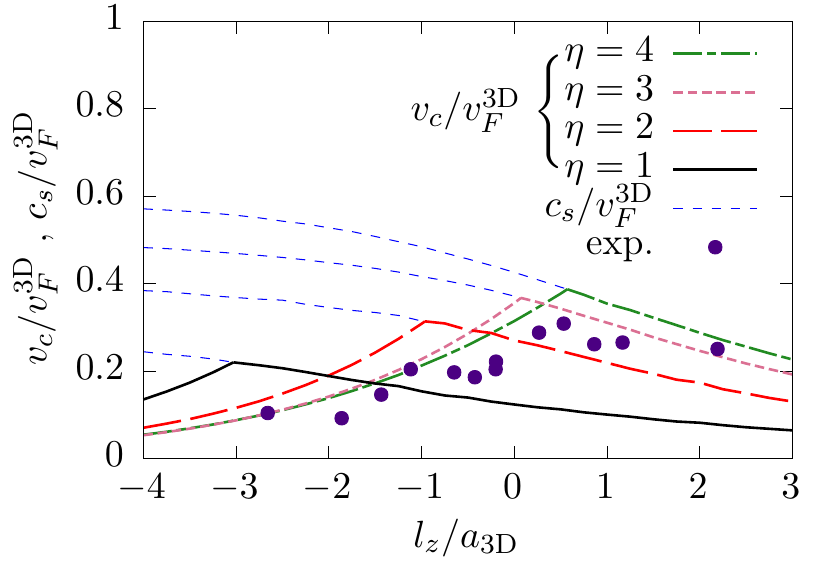} \caption{(color online). Landau critical velocity $v_{c}$ compared with the
speed of sound $c_{s}$ across the BEC-BCS crossover, as a function
of $l_{z}/a_{\text{3D}}$ at different values of $\eta$. The theoretical
values at large $\eta$ may be compared with the experimental results
of the critical velocity, obtained by Weimer \textit{et al.} \cite{Weimer2015}
for a 3D trapped Fermi gas with $\epsilon_{F}^{\text{3D}}/\hbar\omega_{z}\apprge4.2$,
which in our dimensional crossover units roughly corresponds to a
dimension parameter $\eta=k_{F}^{\text{3D}}l_{z}^{\text{HO}}=\sqrt{\hbar\epsilon_{F}^{\text{3D}}/(M\omega_{z})}=2.9$.
Here, $v_{F}^{\textrm{3D}}=\hbar k_{F}^{\textrm{3D}}/M$ is the 3D
Fermi velocity. \label{fig:Landau_critical_velocity}}
\end{figure}

\begin{figure*}
\noindent \centering{}\includegraphics[width=0.32\textwidth]{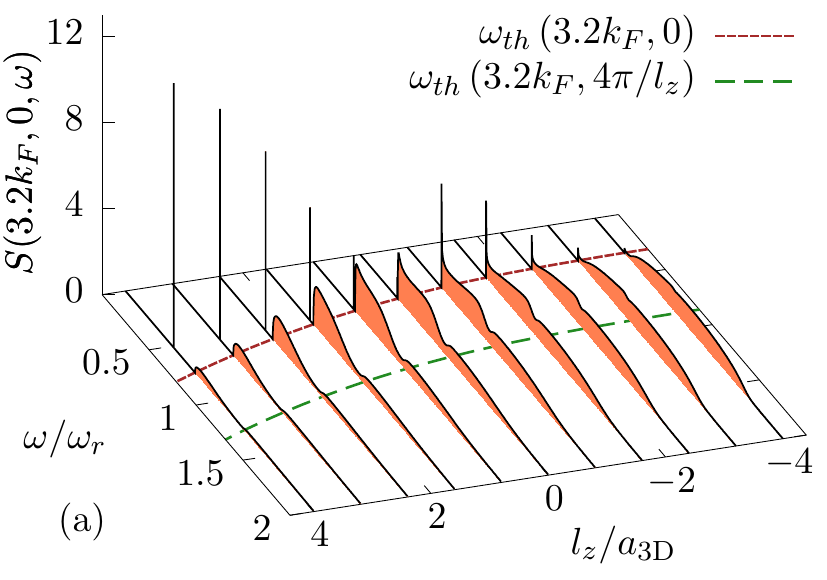}
\includegraphics[width=0.32\textwidth]{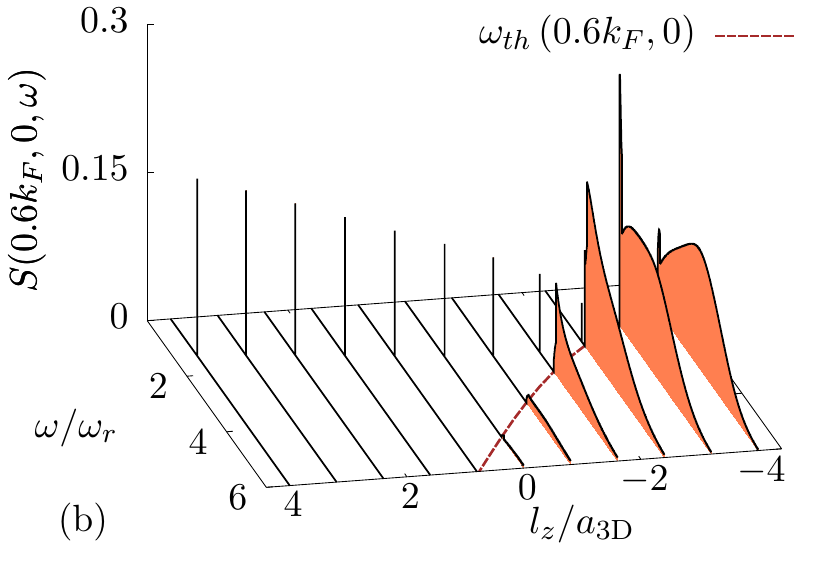} \includegraphics[width=0.32\textwidth]{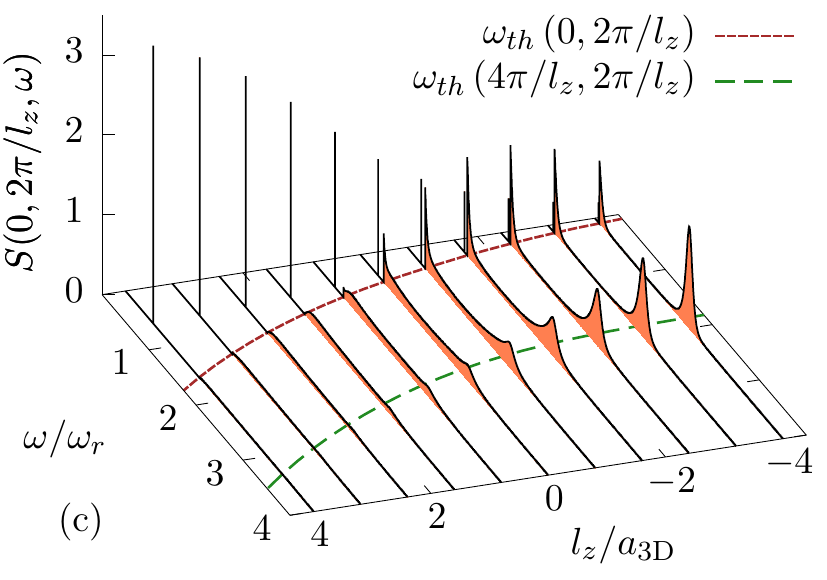}
\caption{(color online). The density dynamic structure factor $S(\mathbf{Q}_{r},\omega)$
scaled by the ratio $\omega_{r}/N$, where $\omega_{r}$ is the recoil
energy and $N$ the particle number, in the quasi-2D regime for $\eta=4$
at various interaction strengths $l_{z}/a_{\text{3D}}$, with the
in-plane recoil momentum $\mathbf{Q}_{r}=(3.2k_{F},0)$ (a), $\mathbf{Q}_{r}=(0.6k_{F},0)$
(b) and transverse recoil momentum $\mathbf{Q}_{r}=(0,2\pi/l_{z})$
(c). The spectral width of the Bogoliubov-Anderson phonon peak is illustrated by the height of the %
delta function. \label{fig:DSF}}
\end{figure*}

\textit{Landau critical velocity.} \textemdash{} Within the GPF theory,
we can calculate the critical velocity of the superfluid
through both the BEC-BCS and dimensional crossover. Once we know the
dispersion of in-plane ($q_{z}=0$) collective modes $\omega_{0}(q=\left|\mathbf{q}\right|)$,
which corresponds to the poles of $[\mathbf{M}_{11}(Q)\mathbf{M}_{11}(-Q)-\mathbf{M}_{12}^{2}(Q)]^{-1}$,
for a given set of the parameters $\eta$ and $l_{z}/a_{\text{3D}}$,
we compute the speed of sound of the superfluid, $c_{s}=\lim_{q\rightarrow0}\omega_{0}(q)/q$,
and the pair-breaking velocity $v_{pb}=[\hbar^{2}(\sqrt{\Delta^{2}+\mu^{2}}-\mu)/M]^{1/2}$
\cite{Combescot2006}. According to Landau's criterion, the critical
velocity in the BEC-BCS crossover is then given by, 
\begin{equation}
v_{c}=\min_{q\geq0}\frac{\omega_{0}\left(q\right)}{q}=\min\left\{ c_{s},v_{pb}\right\} .
\end{equation}

In Fig. \ref{fig:Landau_critical_velocity},
we present the speeds of sound $c_{s}$ and critical velocities
$v_{c}$ for dimensions $\eta=1\sim4$ as a function of the interaction
strength $l_{z}/a_{{\rm 3D}}$. The critical velocity of a 3D Fermi %
superfluid at the BEC-BCS crossover
has been experimentally measured in a harmonic trap~\cite{Miller2007,Weimer2015} %
and can be compared with our results using the transverse
harmonic oscillator length, $l_{z}^{\text{HO}}=\sqrt{\hbar/(m\omega_{z})}$,
as input to determine $\eta$. 
In Ref. \cite{Weimer2015}, the 3D regime is approximately
reached with $\epsilon_{F}^{\text{3D}}/(\hbar\omega_{z})\apprge 4.2$ that corresponds %
to $\eta\simeq2.9$ and the data match qualitatively well with the %
predicted Landau critical velocity (i.e., at $\eta=3$). %

In 3D the BCS regime displays a large speed of
sound and a smaller pair-breaking velocity that limits %
the critical velocity \cite{Combescot2006,Sensarma2006}. On the BEC
side, close to the 3D unitarity, the pair-breaking velocity becomes
equal to the speed of sound, which is referred to as the most robust
configuration of the BEC-BCS crossover \cite{Combescot2006}. Beyond
this point the speed of sound becomes the critical velocity, marking
the system undergoing macroscopic condensation. The 2D and quasi-2D
critical velocities behave similarly to the 3D case. However, the
tuning point of the BEC-BCS crossover $(l_{z}/a_{\text{3D}})_{v_{c}^{\max}}$
\textendash{} at which the critical velocity peaks \textendash{} shows
a non-trivial dependence on the dimensional parameter $\eta$. This
enables us to characterize the dimensional crossover
diagram in the presence of strong interactions, as shown in Fig. \ref{fig:phase_diagram}.
In the region $0\leq\eta<2$, the 2D regime, we see the logarithmic
dependence of the critical velocity maximum with respect to $\eta$, with %
the peak of the critical velocity in 2D at $\ln(k_{F}^{\text{2D}}a_{\text{2D}})\simeq1.08$
\cite{footnote,Shi2015}. Moreover, a linear behavior is observed
in the nearly 3D regime with $\eta>8$ placing the peak of the critical
velocity in 3D at $1/(k_{F}^{\text{3D}}a_{\text{3D}})\simeq 0.056$
\cite{footnote,Combescot2006}. In between ($2<\eta<8,$), the maximum
of the critical velocity lies in the interval $-1<l_{z}/a_{\text{3D}}<0.67$
and $(l_{z}/a_{\text{3D}})_{v_{c}^{\max}}$ varies non-monotonically
with $\eta$. We identify this as the quasi-2D regime, consolidating
the previous conclusion made from equation of state.

\textit{Probing the quasi-2D regime.} \textemdash{} 
A practical way to measure both the speed of sound, $c_s$, and the %
order parameter, $\Delta$, is via Bragg spectroscopy.
The spectroscopic response probes the dynamic structure
factor \cite{Brunello2001,Veeravalli2008,Hoinka2013}, which in the
case of a Fermi superfluid exhibits a peak corresponding to
the Bogoliubov-Anderson phonon mode and a continuum of particle-hole
excitations \cite{Combescot2006}. Due to the presence
of a pairing gap in the excitation spectrum, an external excitation
of momentum $\mathbf{Q}_{r}$ is collective if it does not break pairs
when it excites states with energy below the threshold, 
\begin{equation}
\omega_{\textrm{th}}(\mathbf{Q}_{r})=\left\{ \begin{array}{cc}
2\Delta & \textrm{\ensuremath{\mu>0} and \ensuremath{\hbar^{2}\mathbf{Q}_{r}^{2}\leq8M\mu}}\\
2\sqrt{\mu_{\mathbf{Q}_{r}}^{2}+\Delta^{2}} & \textrm{otherwise}
\end{array}\right.,
\end{equation}
where $\mu_{\mathbf{Q}_{r}}=\mu-\hbar^{2}\mathbf{Q}_{r}^{2}/(2M)$,
and for our dimensional crossover system with finite transverse periodic
length $l_{z}$, we have set $\mathbf{Q}_{r}=(q_{r},q_{z})$, a combination
of an in-plane momentum $q_{r}$, and a transverse excitation, $q_{z}=2\pi n_{z}/l_{z}$
for fixed integer $n_{z}$. We note that the calculation of the dynamic
structure factor $S(\mathbf{Q}_r,\omega)$ within the GPF theory is notoriously difficult \cite{He2016}, %
so we instead use the random phase approximation within the mean-field %
framework \cite{Zou2010}.

In Fig. \ref{fig:DSF}, we plot the dynamic structure factor in the
quasi-2D regime at $\eta=4$, normalized by the number of particles
$N$ and recoil energy $\omega_{r}=\hbar\mathbf{Q}_{r}^{2}/(2M)$
for three different recoil momenta, (a) $\mathbf{Q}_{r}=(3.2k_{F},0)$,
(b) $\mathbf{Q}_{r}=(0.6k_{F},0)$ and (c) $\mathbf{Q}_{r}=(0,2\pi/l_{z})$.
One observes in Figs. \ref{fig:DSF}(a)-(b) that the response is similar
to the 3D case \cite{Zou2010}, showing the characteristic peaks in
the continuum spectrum for $\omega>\omega_{\textrm{th}}$, and %
the presence of the phonon mode. We note the appearance of a
second peak, marked by $\omega_{\textrm{th}}(q_{r},4\pi/l_{z})$ (green
dashed) in Fig. \ref{fig:DSF}(a), corresponding to the generation %
of a transverse excitation. The response at $\omega_{\textrm{th}}(q_{r},2\pi/l_{z})$
is absent, due to the need of the system to excite two modes along
$z$ with opposite momenta, in order to conserve the total momentum.
The same structure, present in Fig. \ref{fig:DSF}(b), is not resolved %
due to the energy required at this momentum. %

The dynamic response of the system, for a transverse recoil momentum
$\mathbf{Q}_{r}=(0,2\pi/l_{z})$, is shown in Fig. \ref{fig:DSF}(c),
and has a specific structure due to the quasi-2D regime. Conservation
of total momentum forces in-plane excitations to place the second
continuum peak at $\omega_{\textrm{th}}(4\pi/l_{z},2\pi/l_{z})$ and
gives no response at $\omega_{\textrm{th}}(2\pi/l_{z},2\pi/l_{z})$,
which would break momentum conservation. We expect this to be a signature
of the quasi-2D regime, as in 3D the pairing gap between
box modes, $2\pi/l_{z}$, goes to zero and the isolated peaks merge
in a continuous structure, while in 2D, the peak $\omega_{\textrm{th}}(4\pi/l_{z},2\pi/l_{z})$
moves too far away from the main spectrum.

\textit{Conclusions}. \textemdash{} In summary, we have examined the
role of dimension in a strongly interacting Fermi superfluid by treating
the transverse confinement with PBC. We have mapped out a dimensional
crossover diagram from the zero-temperature equation of state and
have quantitatively determined the boundaries between 2D, quasi-2D,
and 3D from the location of maximum Landau critical velocity. 
This sets a framework for characterizing the %
BCS-BEC crossover in quasi-2D, %
where the different regimes of the superfluid %
can be experimentally probed using Bragg spectroscopy.  
Our results are directly applicable to an interacting dimensional
crossover Fermi gas realized by imposing a box trapping potential
in the tight confinement direction \cite{Mukherjee2016}, and we expect our findings to %
be qualitatively similar under harmonic transverse confinement. 
\begin{acknowledgments}
This research was supported under Australian Research Council's Discovery
Projects funding scheme (project numbers DP140100637 and DP140103231)
and Future Fellowships funding scheme (project numbers FT130100815
and FT140100003). XJL was supported in part by the National Science
Foundation under Grant No. NSF PHY11-25915, during her visit to KITP.
All numerical calculations were performed using Swinburne new high-performance
computing resources (Green II). 
\end{acknowledgments}

\pagebreak
\widetext
\begin{center}
\textbf{\large Supplemental Material for\\ ``Dimensional crossover in a strongly interacting %
ultracold atomic Fermi gas''}
\end{center}
\setcounter{equation}{0}
\setcounter{figure}{0}
\setcounter{table}{0}
\setcounter{page}{1}
\makeatletter
\renewcommand{\theequation}{S\arabic{equation}}
\renewcommand{\thefigure}{S\arabic{figure}}

\title{Supplemental Material for\\ ``Dimensional crossover in a strongly interacting %
ultracold atomic Fermi gas''}

\author{Umberto Toniolo}

\affiliation{Centre for Quantum and Optical Science, Swinburne University of Technology,
Melbourne 3122, Australia}

\author{Brendan C. Mulkerin}

\affiliation{Centre for Quantum and Optical Science, Swinburne University of Technology,
Melbourne 3122, Australia}

\author{Chris J. Vale}

\affiliation{Centre for Quantum and Optical Science, Swinburne University of Technology,
Melbourne 3122, Australia}

\author{Xia-Ji Liu}

\affiliation{Centre for Quantum and Optical Science, Swinburne University of Technology,
Melbourne 3122, Australia}

\affiliation{Kavli Institute for Theoretical Physics, UC Santa Barbara, USA}

\author{Hui Hu}

\affiliation{Centre for Quantum and Optical Science, Swinburne University of Technology,
Melbourne 3122, Australia}
\date{\today}
\maketitle

\section{Alternative characterizations of the dimensional crossover}

The dimensional crossover is tuned by the quasi-2D BCS-BEC crossover parameter, $l_z/a_{\text{3D}}$, %
computed at the position where the Landau critical velocity has a maximum. Here, we present 
alternative chacterizations by using the ratio between the pairing order parameter $\Delta$ %
and the chemical potential $\mu$, or the ratio between the pairing order parameter and the Fermi energy $\varepsilon_F$.

\begin{figure*}[!h]
\centering
\includegraphics[width=0.45\textwidth]{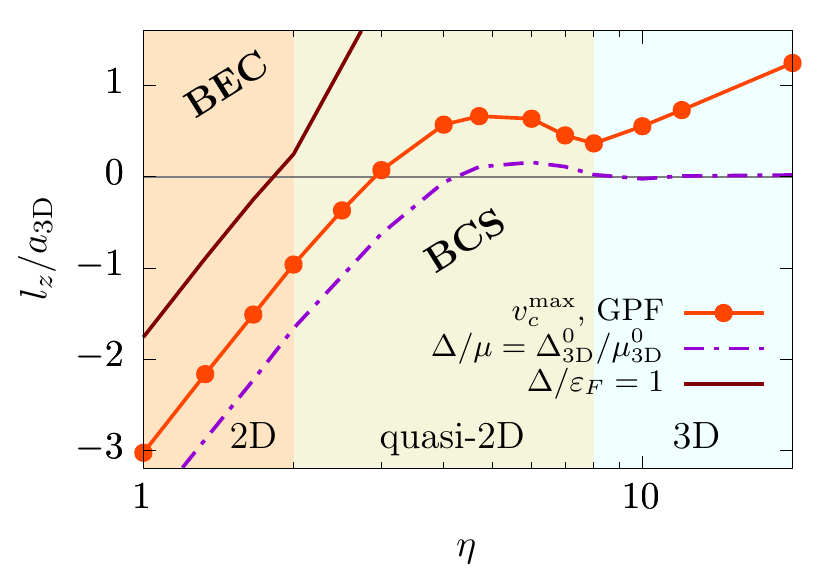}
\caption{\label{fig:supp1}The evolution of the tuning parameter $l_z/a_{\text{3D}}$, computed at different %
conditions, as a function of the dimensional parameter $\eta$ (in logarithmic scale). %
$l_z/a_{\text{3D}}$ is computed (i) when the Landau critical velocity has a maximum (red line with circles), (ii) when the ratio between the order parameter $\Delta$ and the chemical potential $\mu$ reproduces the 3D typical values $\Delta_{\text{3D}}^0=0.46\varepsilon_F^{\text{3D}}$ %
and $\mu_{\text{3D}}^0=0.4\varepsilon_F^{\text{3D}}$~\cite{Hu2006} (purple dot-dashed line), and (iii) when $\Delta=\varepsilon_F$ (solid line at the top right).}
\end{figure*}

In Fig.~\ref{fig:supp1} we plot the critical values of $l_z/a_{\text{3D}}$, across the %
dimensional crossover, when, (i) the Landau critical velocity has a maximum (circles), (ii) the ratio $\Delta/\mu$ 
is equal to the 3D case (dashed-dotted) and (iii) when the order parameter, $\Delta$, is equal to the Fermi energy, $\varepsilon_F$. %
We observe that as expected, the ratio $\Delta/\mu$ approaches the 3D limit for $\eta\rightarrow\infty$, while the %
condition $\Delta=\varepsilon_F$ has meaning only in the far 2D limit. Indeed, the condition $\Delta=\varepsilon_F$ %
can be reached in 3D only at very large values of the tuning parameter $1/(k_F^{\text{3D}}a_{\text{3D}})$, in the %
deep BEC regime. We remark that the choice of the Landau critical velocity, as the most useful condition to %
characterize the crossover, allows a complete independent description %
from both the 3D and 2D regimes, since the interaction effect is fully taken into account in
$v_c^{\max}$.

\section{Landau critical velocity in the 3D and 2D limits}

\begin{figure*}
\centering
\includegraphics[width=0.45\textwidth]{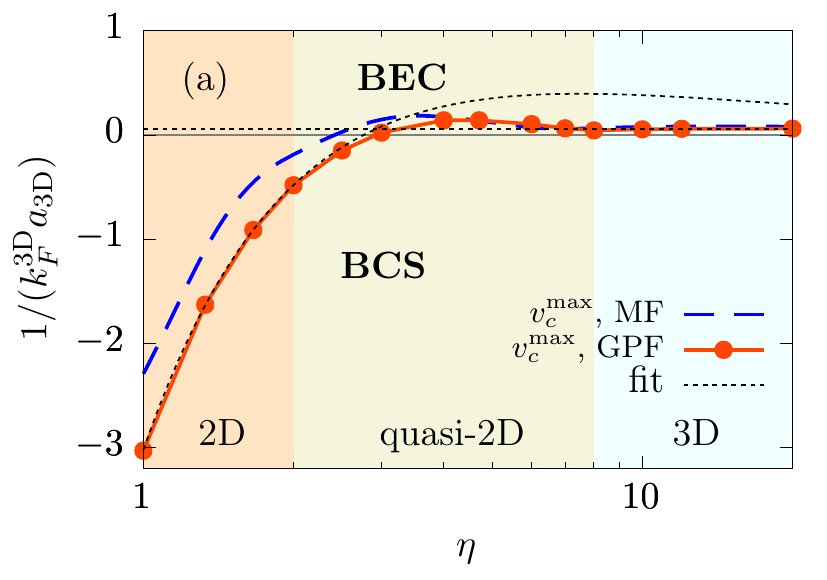}
\includegraphics[width=0.45\textwidth]{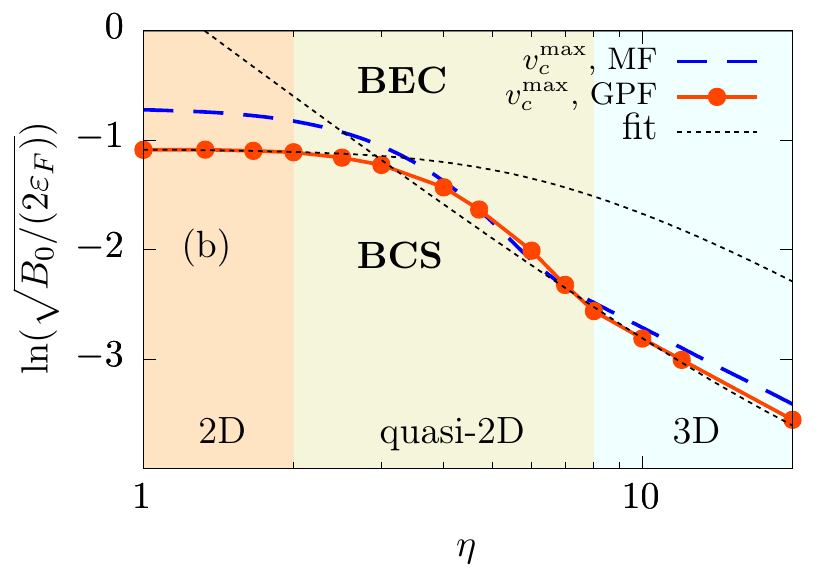}
\caption{\label{fig:supp2}The dimensional crossover diagram tuned by the dimensional parameter, $\eta$, and %
(a) the typical 3D BCS-BEC %
crossover parameter, $1/(k_F^{\text{3D}}a_{\text{3D}})$, and (b) the typical 2D BCS-BEC %
crossover parameter, $\ln(\sqrt{B_0/(2\varepsilon_F)})$. The fitted lines (tiny dotted) %
show (a) the expected constant behaviour $[1/(k_F^{\text{3D}}a_{\text{3D}})]_{v^{\max}_c}\simeq 0.056$ %
at $\eta\rightarrow\infty$, and (b) the expected constant behaviour %
$[\ln(\sqrt{B_0/(2\varepsilon_F)})]_{v_c^{\max}}\simeq -1.08$ at $\eta\rightarrow 0$.}
\end{figure*}

\subsection{The 3D limit}
We observe that, for a 3D Fermi gas, %
the MF theory predicts the critical velocity to be slightly on the BEC side at 
approximately $1/(k_F^{\text{3D}}a_{\text{3D}})_{v^{\max}_c}\simeq 0.07$~\cite{Combescot2006}. Figure~\ref{fig:phase_diagram} is expected %
to predict this behaviour when we restore the 3D limit case for $\eta\rightarrow\infty$. We 
consider the most general choice to descibe the BCS-BEC tuning parameter in this limit,  
\begin{equation}
\left(\frac{l_z}{a_{\text{3D}}}\right)_{v_c^{\max}}=\sum_{n=-\infty}^{\infty}a_n\eta^n.
\end{equation}
The limit of the proper 3D tuning parameter, $1/(k_F^{\text{3D}}a_{\text{3D}})$, is then given by 
\begin{equation}
\lim_{\eta\rightarrow\infty}\left(\frac{l_z}{a_{\text{3D}}}\right)_{v_c^{\max}}\frac{1}{\eta}%
=\sum_{n=-\infty}^{\infty}a_n\eta^{n-1}.
\end{equation}
Every coefficient $a_n$ for $n<1$ is negligible when $\eta$ is large enough, %
while each term $a_n\neq 0$, for $n>1$, would lead to a divergence in the definition of the 3D peak %
for the critical velocity and it is therefore discarded. We fit the far right hand side of %
Fig.~\ref{fig:phase_diagram} via a linear function, 
\begin{equation}
\left(\frac{l_z}{a_{\text{3D}}}\right)_{v_c^{\max}}\Big|_{\eta\geq 8}=a_0+a_1\eta,
\end{equation}
and we included the $a_0$ term due to the proximity of data to the quasi-2D regime when $\eta\simeq 8$. This leads to%
\begin{equation}
\left(\frac{1}{k_F^{\text{3D}}a_{\text{3D}}}\right)_{v^{\max}_c}=a_1\simeq 0.056.
\end{equation}
The behaviour in the 3D regime is shown in Fig.~\ref{fig:supp2}(a) that provides the same results of Fig.~\ref{fig:phase_diagram} %
with a change of scale in the vertical axis from $l_z/a_{\text{3D}}$ to $1/(k_F^{\text{3D}}a_{\text{3D}})$. 

\subsection{The 2D limit}
In the 2D limit, we denote that 
\begin{equation}
\left(\frac{l_z}{a_{\text{3D}}}\right)_{v_c^{\max}}=\mathcal{F}(\eta),
\end{equation}
where the approximate form of  $\mathcal{F}(\eta)$ is to be determined. We observe from Fig.~\ref{fig:EoS}(a)-(b) that the proper BCS-BEC crossover tuning parameter becomes $\ln[\sqrt{B_0/(2\varepsilon_F)}]$, where $k_F=k_F^{\text{2D}}$, for $\eta\leq\pi 6^{1/3}$, and %
$\lim_{\eta\rightarrow 0}\sqrt{B_0/(2\varepsilon_F)}=1/(k_F^{\text{2D}}a_{\text{2D}})$. We consider the relation between the %
2D and the quasi-2D tuning parameters, 
\begin{equation}
\sqrt{\frac{B_0}{2\epsilon_F}}=\frac{\sqrt{3\pi}}{\eta^{3/2}}\text{arcsinh}\left[%
\frac{1}{2}\exp\left(\frac{l_z}{2a_\text{3D}}\right)\right]. 
\end{equation}
It is reasonable to assume that at the position where the Landau critical velocity takes the maximum value, we would have,
\begin{equation}
\lim_{\eta\rightarrow 0}\sqrt{\frac{B_0}{2\epsilon_F}}=A\neq 0.
\end{equation}
By using the above three equations, we find that,
\begin{equation}
\mathcal{F}(\eta)=2\ln\left[\sinh\left(\frac{A\eta^{3/2}}{\sqrt{3\pi}}\right)\right]%
\sim_{\eta=0}2\ln\left(\frac{2A}{\sqrt{3\pi}}\right)+3\ln\eta+\mathcal{O}(\eta^3).
\end{equation} 
Therefore, our numerical results in the 2D limit could be fitted with the function  %
\begin{equation}
\mathcal{F}(\eta)=W+Z\ln{\eta}.
\end{equation}
We obtain the values $W=3.02$ and $Z=2.97$ from the fitting. The latter value $Z\simeq 3$ confirms our theoretical anticipation of the 2D limit, within a relative error of a few percents. Using the former value of $W$, we can compute the position of the Landau critical velocity %
peak in the 2D limit being, 
\begin{equation}
\lim_{\eta\rightarrow 0}\ln\left(\sqrt{\frac{B_0}{2\epsilon_F}}\right)_{v_c^{\max}}=\ln A=\frac{W}{2}+\ln\frac{\sqrt{3\pi}}{2}\simeq -1.08,
\end{equation}
as represented in the dimensional crossover diagram of Fig.~\ref{fig:supp2}(b) %
that provides the same results of Fig.~\ref{fig:phase_diagram} %
with a change of scale in the vertical axis from $l_z/a_{\text{3D}}$ to $\ln[\sqrt{B_0/(2\varepsilon_F)}]$. 
This extracted position of the peak of the Landau critical velocity in the 2D limit is consistent with the position %
of the peak of the contact, $\ln(k_F^{2D}a_{2D})\sim1$, obtained recently via auxiliary-field Monte Carlo simulations for a 2D interacting Fermi gas~\cite{Shi2015}. 
\end{document}